\documentclass[aps,prb,twocolumn,groupedaddress,showpacs]{revtex4-1}
\usepackage[english]{babel}
\usepackage{stmaryrd}
\usepackage{amssymb}
\usepackage{amsfonts}
\usepackage{amsmath}
\usepackage{color}
\usepackage{xspace}
\usepackage{graphicx}
\usepackage{bm}

\begin{document}

\title{Nature of the high-speed rupture of the two-dimensional Burridge-Knopoff model of earthquakes}

\author{Hikaru Kawamura}
\email{kawamura@ess.sci.osaka-u.ac.jp}

\author{Koji Yoshimura}
\author{Shingo Kakui}
\affiliation{Graduate School of Science, Osaka University,
Toyonaka, Osaka 560-0043, Japan}

\date{\today}

\begin{abstract}
The nature of the high-speed rupture or the main shock of the Burridge-Knopoff spring-block model in two dimensions obeying the rate-and-state dependent friction law is studied by means of extensive computer simulations. It is found that the rupture propagation in larger events is highly anisotropic and irregular in shape on longer length scales, although the model is completely uniform and the rupture-propagation velocity is kept constant everywhere at the rupture front. The manner of the rupture propagation sometimes mimics the successive ruptures of neighboring ``asperities'' observed in real large earthquakes. Large events tend to be unilateral, with its epicenter lying at the rim of its rupture zone. The epicenter site is also located next to the rim of the rupture zone of some past event. Event-size distributions are computed and discussed in comparison with those of the corresponding one-dimensional model. The magnitude distribution exhibits a power-law behavior resembling the Gutenberg-Richter law for smaller magnitudes, which changes over to a more characteristic behavior for larger magnitudes. The behavior of the rupture length for larger events is discussed in terms of the strongly anisotropic rupture propagation of large events reflecting the underlying geometry.
\end{abstract}

\maketitle


\section{Introduction}

 Earthquake may be regarded as a stick-slip dynamical instability of a fault driven by the tectonic plate motion \cite{Scholz2002,Kanamori}. While an earthquake is obviously a complex large-scale dynamical instability involving large degree of freedom, general expectation would be that it should be describable by a set of classical equations of motion with appropriate friction laws or constitutive relations \cite{Scholz1998,Scholz2002}. 

  Statistical physical study of earthquakes is often based on simplified models of various levels of simplification \cite{Kawamura2012}. One of the standard models widely employed in statistical physical study of earthquakes might be the Burridge-Knopoff (BK) spring-block model \cite{CLS,Kawamura2012}. The model was first introduced in Ref.\cite{BK}. In the BK model, an earthquake fault is simulated by an assembly of blocks, each of which is connected via the elastic springs to the neighboring blocks and to the moving plate.

 The properties of the BK model have extensively been studied for years now. At an earlier stage of the study, Carlson, Langer and collaborators performed a series of numerical simulations on the model, paying particular attention to the magnitude distribution of earthquake events  \cite{CL1989a,CL1989b,CLST,Carlson1991a,Carlson1991b,Shaw1992,CLS}. In these simulations, as a friction law, a simplified friction law, a velocity-weakening law, was employed where the friction force is assumed to be a single-valued decreasing function of the sliding velocity. These earlier simulations were further extended in several ways, computing various observables\cite{Schmittbuhl,MoriKawamura2005,MoriKawamura2006}, extending the model to two dimensions \cite{Carlson1991b,Myers,MoriKawamura2008a,MoriKawamura2008b}, taking account of the effect of the viscosity\cite{ML,Shaw1994,De,MoriKawamura2008c}, modifying the form of the friction force \cite{ML,Shaw1995,Cartwright,De,Clancy}, considering the long-range interactions between blocks \cite{Xia,MoriKawamura2008b}, driving the system only at one end of the system (the train model) \cite{Vieira}, and examining the continuum limit \cite{MoriKawamura2008c}, {\it etc\/}.

 More realistic constitutive relation now standard in seismology might be the rate-and-state dependent friction (RSF) law \cite{Scholz1998,Scholz2002,Dieterich, Ruina, Marone}. The RSF law assumes that the friction depends not only on the slip velocity but also on the ``state'' of the slip interface, which is phenomenologically described via the ``state variable'' obeying its own evolution law. The time-evolution law of the state variable generally includes a characteristic slip distance ${\mathcal L}$, which gives a measure of the length scale at which a slip interface loses its initial memory of the state. Then, the RSF law describes the friction as a  competition between the two effects, one is a direct velocity-strengthening friction represented by the so-called `$a$-term', and the other is an indirect velocity-weakening friction represented by the  so-called `$b$-term'.
 
 In this situation, it would be highly interesting and desirable to clarify the properties of the BK model obeying the RSF law \cite{CaoAki}. Systematic studies on the statistical properties of the 1D BK model under the RSF law have been performed by the present authors' group. Ohmura and Kawamura \cite{OhmuraKawamura}, and subsequently Kawamura {\it et al\/} \cite{Kawamura2017} studied the statistical properties of the model, and revealed that they differed significantly from those of the BK model under the simple velocity-weakening law. Namely, under the RSF law, earthquake events exhibit more enhanced ``characteristic'' features, with characteristic energy, length and time scales. For example, the magnitude distribution often largely deviates from the scale-invariant power-law (Gutenberg-Richter law) behavior. Such a deviation was also observed for the BK model under the simple velocity-weakening friction law, but more so in the BK model under the RSF law.

  Recent studies by Ueda {\it et al\/} revealed that the BK model under the RSF law could describe a slow nucleation process preceding the high-speed rupture of a main shock \cite{Ueda2014,Ueda2015}, which is not describable by the BK model under the simple velocity-weakening law. Furthermore, recent studies by Kawamura, Yamamoto and Ueda have revealed that the BK model under the RSF law can also describe the slow-slip phenomena including afterslips and silent earthquakes \cite{Kawamura2018}. Afterslips are long-lasting slow-slip process which take place following the  high-speed rupture of main shock, while silent earthquakes are slow-slip-only process not accompanying any high-speed rupture nor emission of seismic waves. In fact, recent development in GPS technology and in high-density GPS and seismograph networks has revealed these slow-slip phenomena where the fault sliding velocity is several orders of magnitudes slower than that of the standard high-speed rupture \cite{Kawasaki1995,Heki,Hirose,Dragert,Obara,Miller,Ozawa,Kostoglodov,Kawasaki2004}. While such slow-slip phenomena have turned out to be well reproducible by the BK model under the RSF law, but are not describable by the BK model under the simple velocity-weakening law. In this way, the 1D BK model under the RSF law allows reproducing a variety of seismic slips within a single framework, including main shocks, precursory nucleation processes, afterslips, and silent earthquakes by varying only a few model parameters. 

 So far, these studies on the BK model under the RSF law have been made for the 1D model. Under such circumstances, it is desirable to investigate the properties of the corresponding 2D model. In the present paper, we try such simulations on the 2D BK model under the RSF law, by concentrating on the high-speed rupture of main shocks. Slow slips and nucleation processes are not considered here, and will be left for future works. We will concentrate on the parameter region of the strong frictional instability regime where the simulation is the easiest, but instead, try to simulate sufficiently large system to be free from finite-size effects, and try to generate sufficiently many events to get good statistics.

 This paper is organized as follows. In section II, we introduce our model, the 2D BK model obeying the RSF law, and present its equation of motion. Some of the details of the simulation is explained here. Section III is the main part of this paper, where we report the results of our numerical simulations on various properties of the model, {\it e.g.\/}, the way of the rupture propagation, the magnitude distribution, the rupture-length distribution, the mainshock recurrence-time distribution, {\it etc\/}. Section IV is devoted to summary and discussion.

\section{The model}

 The 2D BK model obeying the RSF law considered here consists of a 2D array of $N=L\times L$ identical blocks with the mass $m$, which are mutually connected by four neighboring blocks via elastic springs with spring stiffness $k_c$, and these are connected to a moving plate via springs with spring stiffness $k_p$, as illustrated in Fig.1. All blocks are subject to a friction force $\Phi$. The dimensionless equation of motion for the block at the site ($i,j$) can be written as \cite{CLS,Kawamura2012}

\begin{equation}
\begin{array}{ll}
\ddot u_{i,j}=\nu t-u_{i,j}+l^2(u_{i+1,j}+u_{i,j+1} \ \ \ \ \ \\
\ \ \ \ \  +u_{i-1,j}+u_{i,j-1}-4u_{i,j})-\phi_{i,j} ,
\end{array}
\end{equation}
where $\ell \equiv (k_c/k_p)^{1/2}$. The dimensionless displacement $u_{i,j}$ is normalized by the characteristic slip distance ${\mathcal L}$ associated with the RSF law, the time $t$ by $\omega^{-1}=\sqrt{m/k_p}$, the block velocity $v_{i,j}$ and the pulling speed of the plate $\nu$ by $\mathcal{L}\omega$, and the dimensionless friction force $\phi_{i,j}$ by $k_p{\mathcal L}$. Note that it is assumed here that the displacement of each block occurs only along the direction of the plate drive. The motion perpendicular to the plate motion is neglected \cite{Carlson1991b}. 

\begin{figure}[ht]
\begin{center}
\includegraphics[scale=0.4]{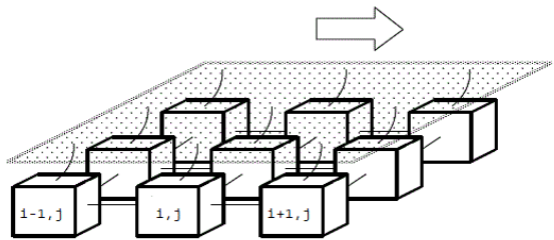}
\end{center}
\caption{
Illustration of the two-dimensional Burridge-Knopoff model.
}
\end{figure}

 The RSF force $\phi_{i,j}$ assumed here reads as \cite{Ueda2014,Ueda2015,Kawamura2017}
\begin{equation}
\phi_{i,j}=c+a\log(1+\frac{v_{i,j}}{v^*})+b\log \theta_{i,j} ,
\label{RSF}
\end{equation}
where $\theta_{i,j}$ is the dimensionless state variable describing the ``state'' of the interface, and the normalized friction parameters $a$, $b$, and $c$ represent velocity-strengthening, velocity-weakening, and constant parts of friction, respectively. The original friction parameters $A$, $B$, and $C$ are related to the normalized parameters by $A=(k_p{\mathcal L}/{\mathcal N})a$, $B=(k_p{\mathcal L}/{\mathcal N})b$, and $C=(k_p{\mathcal L}/{\mathcal N})c$, where ${\mathcal N}$ is the normal load. The state variable $\theta_{i,j}$ is assumed to obey the aging law \cite{Ruina},
\begin{equation}
\frac{{\rm d}\theta_{i,j}}{{\rm d}t} = 1-v_{i,j}\theta_{i,j} .
\label{aging}
\end{equation}

Note that, in contrast to the standard RSF law with the $a$-term being proportional to a pure logarithmic form $\log v$, which yields a pathological limit of a negatively divergent friction for $v\rightarrow 0$, we phenomenologically introduce a modified form using the crossover velocity $v^*$  \cite{Ueda2014,Ueda2015,Kawamura2017}. The modified form, where the $a$ term reduces to a purely logarithmic form when $v>>v^*$ but becomes proportional to the block velocity $v$ when $v<<v^*$, is able to describe a complete halt, unlike the standard form. 

 The model is characterized by only a few basic dimensionless parameters, {\it e.g.\/}, the frictional parameters $a$, $b$, and $v^*$ ($c$ is actually irrelevant \cite{Kawamura2017}), and the elastic parameter $\ell$. The plate velocity $\nu$ is also unimportant so long as it remains small, aside from setting the interseismic timescale. 

 Estimates of typical values of the model units representing natural faults have been given \cite{Ueda2014}. The BK model possesses a built-in time scale $\omega^{-1}$ corresponding to the rise time of an earthquake event. This may be estimated to be $\sim 1$ [s]. The model possesses two distinct and independent length scales: one associated with the fault slip and the other with the distance along the fault. The former length scale is the critical slip distance ${\mathcal L}$, which was estimated to be $\simeq 1$ [cm], while the other is the distance the rupture propagates per unit time, $v_s/\omega$, which was estimated to be $\simeq 2-3$ [km], where $v_s$ is the $s$-wave velocity along the fault. The typical plate velocity is several [cm/year] and corresponds to a very small number of $\nu\sim 10^{-7}-10^{-8}$. The spring constant $k_p$ was related to the normal stress as $\frac{{\mathcal N}}{k_p{\mathcal L}}\simeq 10^2-10^3$. Then, as $C$ is known to take a value around $\frac{2}{3}$, $c$ would be of order $10^2$-$10^3$, with $a$ and $b$ being one or two orders of magnitude smaller than $c$. The crossover velocity $v^*$ is hard to estimate, though it should be much smaller than unity, and we take it as a parameter.

 In our simulations, for the sake of computational feasibility, we concentrate on the strong frictional instability regime where the $b$ parameter is greater than a critical value $b_c$. In fact, $b_c$ is determined solely by the elastic parameter $\ell$. In 1D, it is given by  $b_c=2\ell^2+1$, while in 2D it is given by $b_c=4\ell^2+1$. In the following, we put $\ell=3$ corresponding to $b_c=37$, and vary $b$ in the range of $b>b_c$, {\it i.e.\/}, $40\leq b\leq 80$. Other parameters are fixed to be $a=1$, $c=1000$, $v^*=10^{-2}$ and $\nu=10^{-8}$. 

 Concerning the system size $L$, we put mainly $L=500$, but additionally put $L=1000$ and 1500 to check the system-size dependence. Periodic boundary conditions are assumed in both directions. Concerning the initial conditions, all blocks are assumed to be at rest, {\it i.e.\/}, $v_{i,j}=0$ ($1\leq i,j\leq L$) at $t=0$, and the state variable is set to $v_{i,j}=10^{12}$. The displacement of each block is taken to be random uniformly distributed between [-500, 500] from block to block. Events at earlier times are transient and are affected in a non-stationary manner by the initial conditions. We wait until the system reaches the stationary state, typically discarding initial $10^5$ events. Averages are then taken by using subsequent $10^5 - 10^6$ events.

\section{Results}

 In this section, we present the results of our simulations on the 2D BK model obeying the RSF law. In the stationary state, the model exhibits a sequence of earthquake-like events of various sizes, from small events of only one or two blocks to large events involving, say, $\sim 10^5$ blocks. Among these events, we show in Fig.2(a)-(f) the way of their rupture propagation for six typical large events in the stationary state of the $L=500$ system as color plots of the block sliding velocity. The shot is drawn with the interval time of $\Delta t=10$ (remember that our unit time roughly corresponds to one second in real time). As can be seen from these figures, the way of rupture propagation is neither symmetric nor circular, but is highly asymmetric and irregular in shape. In fact, it sometimes happens that the rupture propagation almost stops at a certain stage, while the rupture propagation resumes again from a certain point on the rim of the (almost) stopped rupture front as if that point were the second epicenter.
\begin{figure*}[ht]
\begin{center}
\includegraphics[width=13cm]{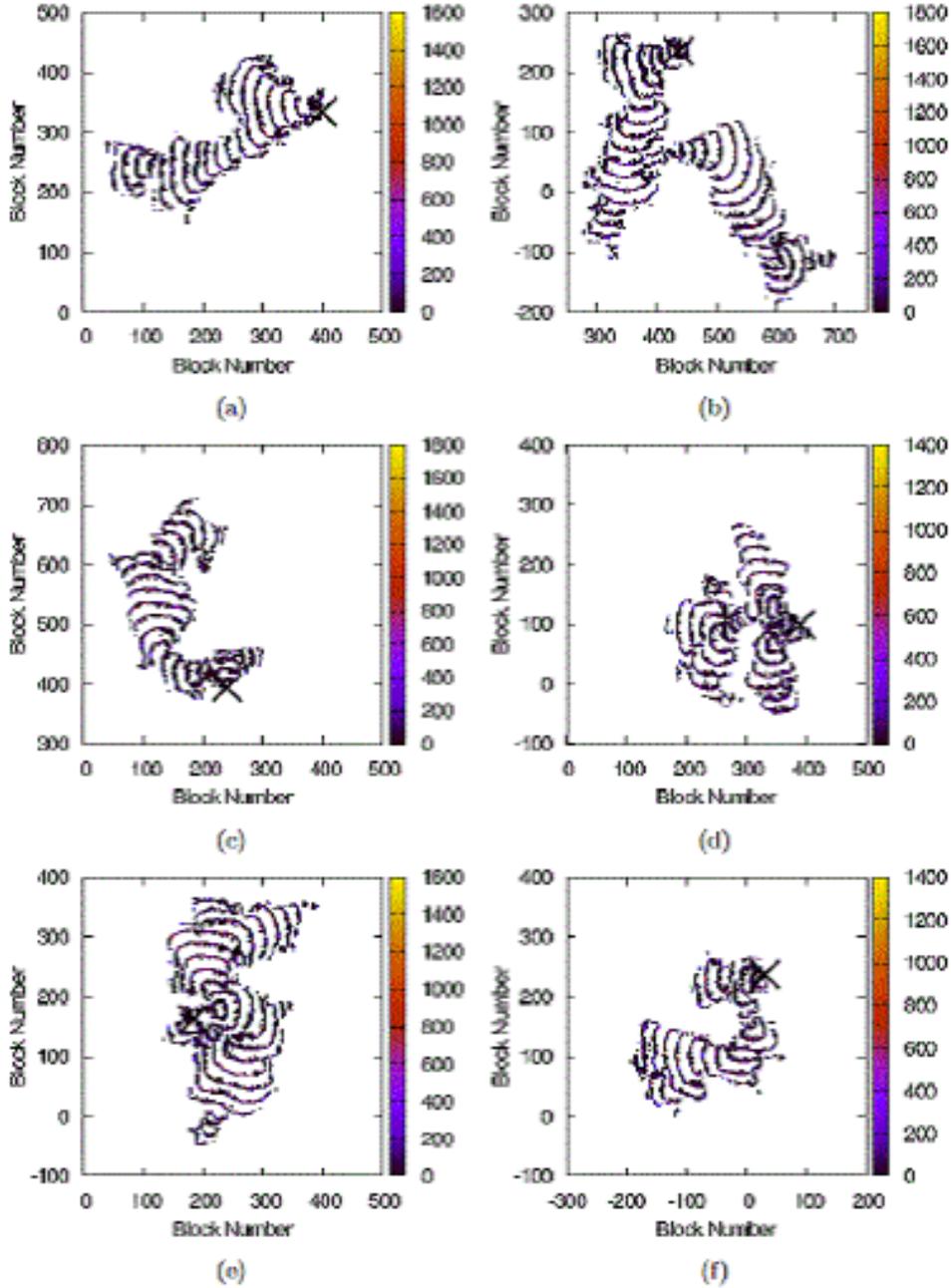}
\end{center}
\caption{
The manner of the rupture propagation for typical large events of the 2D BK model under the RSF law, where the block sliding velocity is shown by the color plot with the time interval of $\Delta t=10$. Figures (a)-(f) represent six typical large events occurring in the stationary state. The cross symbol denotes the epicenter block.  The lattice size is $L=500$. The model parameters are $a=1$, $b=60$, $c=1000$, $\ell=3$, $v^*=10^{-2}$ and $\nu=10^{-8}$.
}
\end{figure*}

 Such a sequence of several sub-events forming a large one event reminds us the successive ruptures of several ``asperities'', sometimes observed in real large earthquakes, where the rupture of one asperity appears to trigger the rupture of the neighboring one. The standard view of the asperity might be that the fault is spatially inhomogeneous in its material parameters, each asperity being more or less rigid as a single body, with a weak link to the neighboring asperities. In the present model, however, the origin of the apparent successuve ruptures of ``asperities'', or more precisely, ``asperity-looking regions'', is very different from such a view. Namely, all the material parameters and the dynamical evolution rule (the equations of motion) are completely uniform, and the strongly anisotropic, asperity-looking rupture propagation and the underlying inhomogeneous stress state are dynamically self-generated from such completely uniform equations of motion and material parameters.

 One should also notice that the rupture propagation velocity $v_r$ itself is kept nearly constant $v_r\sim \ell$ everywhere so long as the rupture propagates. So, the strong anisotropy and irregularity of the rupture propagation is not due to the random modulation of the rupture-propagation velocity.

\begin{figure}[ht]
\begin{center}
\includegraphics[width=8.5cm]{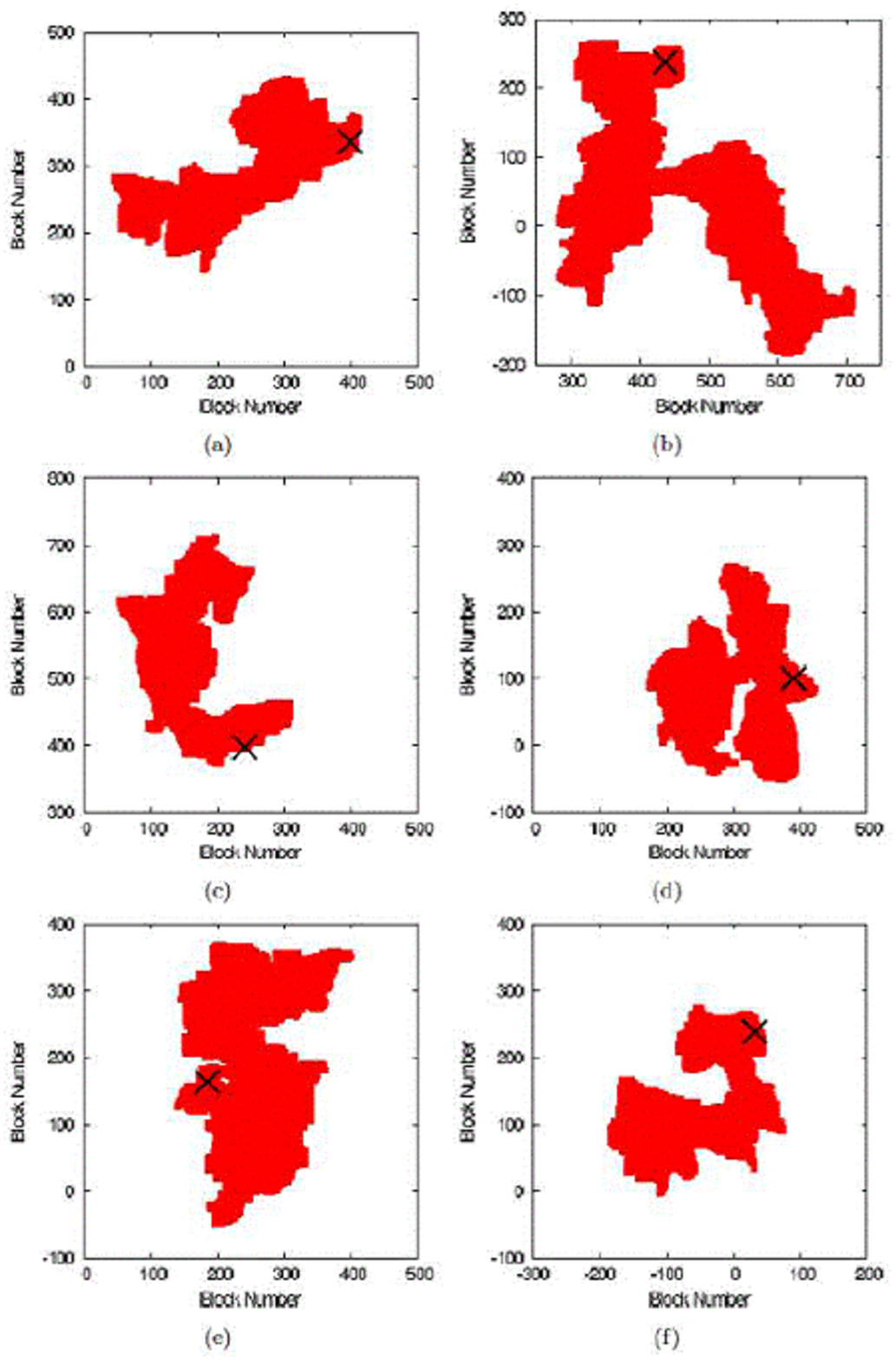}
\end{center}
\caption{
The rupture zone and the epicenter site (indicated by the cross symbol) of the six large events in the stationary state of the 2D BK model under the RSF law shown in Fig.2. Figures (a)-(f) correspond to (a)-(f) of Fig.2, respectively. The lattice size is $L=500$. The model parameters are $a=1$, $b=60$, $c=1000$, $\ell=3$, $v^*=10^{-2}$ and $\nu=10^{-8}$.
}
\end{figure}

 Anyway, such anisotropic and irregular rupture propagation becomes ever possible in 2D, never possible in 1D. Namely, the nontrivial geometry of 2D makes the self-generation of the spatially complex asperity-like seismic structure possible from the completely uniform dynamical rule and material parameters.

 To further clarify the shape of the rupture zone and the epicenter location with respect to the rupture zone, we show in Fig.3 the rupture zone and the epicenter site indicated by the cross symbol for the six large events shown in Fig.2.

 As can be seen from these figures, the epicenter (represented by the cross symbol) is located at the rim of the rupture zone of the event. In other words, the event tends to be unilateral, {\it i.e.\/}, tends to propagate from the epicenter in a biased direction, not propagating isotropically in every direction. In this sense, large events of the present 2D model are `type-I' events as defined in Ref.\cite{Kawamura2017} studying the corresponding 1D model. There, the `type-I' event was defined as a unilateral event with its epicenter lying one block next to the rim of the rupture zone of some past event. Indeed, we confirm that in our 2D model the epicenter of large events is located at the rim of the rupture zone of past events and that most of large events, say, more than 99\% of events, are type-I events.

 In 1D, there also exist type-II events where the epicenter is located in the interior of the rupture zone of that event with its rupture zone possessing a considerable overlap with that of the past event. In the present 2D model, by contrast, we find that such type-II events rarely occur, less than 1\% in number.

 Next, we investigate the event-size distribution. The magnitude of an event, $\mu$, is defined here by
\begin{equation}
\mu = \ln \left( \sum_{i,j} \Delta u_{i,j} \right),
\end{equation}
where the sum is taken over all blocks involved in the event. In Figs.4, we show the magnitude distribution of events in the strong frictional instability regime.
\begin{figure}[ht]
\begin{center}
\includegraphics[width=7cm]{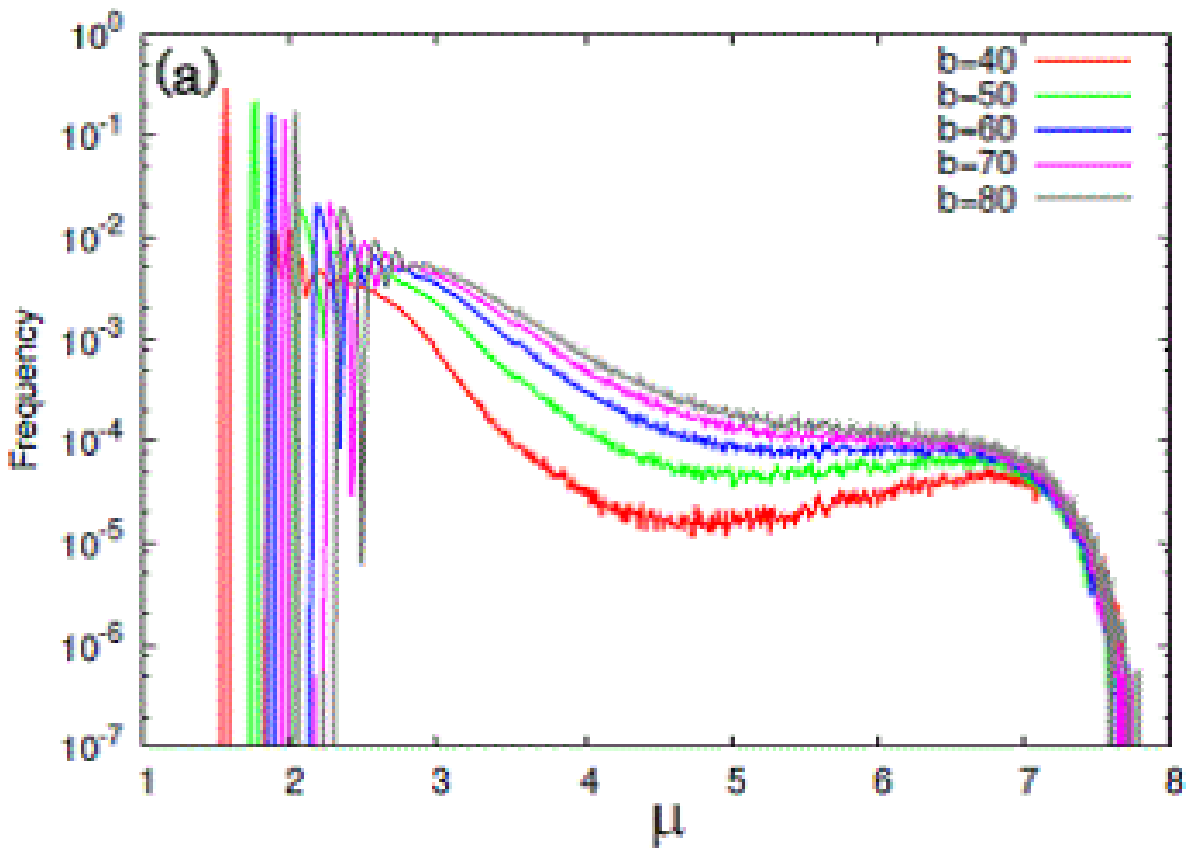}
\includegraphics[width=7cm]{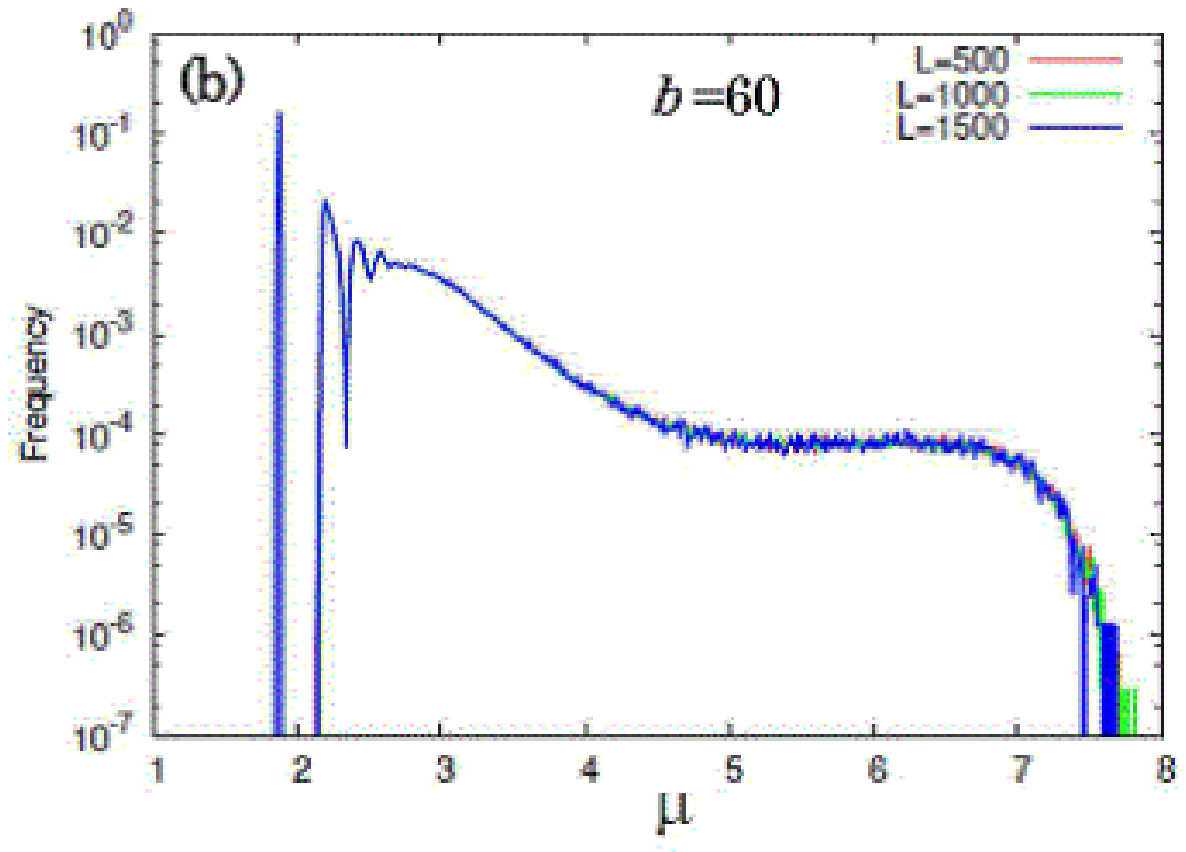}
\end{center}
\caption{
The magnitude distribution of earthquake events of the 2D BK model under the RSF law in the strong frictional instability regime: (a) for various $b$-values with fixing the lattice size to $L=500$; and (b) for various lattice sizes $L$ with fixing the $b$ parameter to $b=60$. The other parameters are $a=1$, $l=3$, $c=1000$, $v^*=10^{-2}$ and $\nu=10^{-8}$.
}
\end{figure}

 As can be seen from Fig.4(a), the computed magnitude distribution exhibits different behaviors for different magnitude regions. For smaller events of $\mu \lesssim 4\sim 5$, the data exhibit near-linear behaviors corresponding to the power law, {\it i.e.\/}, behaviors close to the Gutenberg-Richter (GR) law is observed. The power-law exponent $B$ decreases from $\simeq 1.7$ for $b=40$ to $\simeq 0.9$ for $b=80$.

 For comparison, we show in Fig.5(a) the magnitude distribution of the corresponding 1D model in the strong frictional instability regime in the same range of $40\leq b\leq 80$ as that of the 2D model. The power-law-like behaviors observed at smaller magnitudes in 2D are not visible in 1D.

 For events of medium size of $4 \lesssim \mu \lesssim 7$, the magnitude distribution becomes almost flat, similarly to that of the 1D model of Fig.5(a). For larger events of $\mu \gtrsim 7$, the magnitude distribution exhibits a rapid fall-off with a faster-than power-law decay. At around $\mu \simeq 7$, a shoulder or a weak peak is observed, which tends to be more pronounced as $b$ gets smaller toward the borderline value $b_c=37$. Then, the computed magnitude distribution indicates that, while events at smaller magnitudes exhibit certain `critical' power-law features, events at larger magnitudes exhibit much more `characteristic' features.  

 In order to be sure that such a peak and a rapid fall-off behaviors observed at larger magnitudes are not finite-size effects, we examine the size $L$-dependence.  In Fig.4(b), we show the magnitude distribution on increasing the lattice size as $L=500$, 1000 and 1500 for the representative case of  $b=60$. As can be seen from the figure, the computed magnitude distribution do not show any apprecialbe size effect, indicating that the characteristic feature observed for larger events is an intrinsic bulk effect.

 The magnitude $\mu$ has been defined as the logarithm of the multiple of the mean slip amount and the total number of moving blocks $N_r$ involved in the event proportional to the rupture-zone area, and. One may define the `rupture-length' $L_r$ as a square root of the rupture-zone area, $L_r=\sqrt{N_r}$. In Fig.6(a), we show the distribution of the rupture-length $L_r$ defined in this way on a semi-logarithmic plot for various values of $b$ in the strong frictional instability regime. The parameter choice is the same as in Fig.4(a). The rupture-length distribution exhibits a sharp drop for smaller events of $L_r \lesssim 20-30$ which we call the region I, whereas for intermediate-size events of $20-30 \lesssim L_r \lesssim 170-200$, the distribution exhibits a near stright-line behavior corresponding to an exponential distribution $\simeq \exp[-(L_r/L_0)]$ ($L_0$ is a characteristic rupture length), which we call region II. For larger events of $L_r \gtrsim 170-200$, a faster-than-exponential rapid decay is observed in the frequency distribution, which we call region III. 

 For comparison, we show Fig.5(b) the rupture-length $L_r$ distribution of the 1D model covering the larger-$b$ region than the one studied in Ref.\cite{Kawamura2017}, which just corresponds to the magnitude distribution shown in Fig.5(a). The distribution in 1D exhibits an exponential behavior for a wider range of intermediate and larger events. In particular, such an exponential behavior seems to persist up to the maximum size. In that sense, the region III of the 2D model seems to be absent in 1D.

 Related to this observation, the maximum rupture-length seems to be considerably longer in 1D than in 2D. If one operationally define the maximum rupture-length $L_{r, max}$ here by the $L_r$-value whose frequency is of order $10^{-6}$ to the maximum frequency, it comes around $L_{r,max}\sim 300-400$ in 2D, whereas it is around $L_{r,max}\simeq 1000-1500$ in 1D. The reason of this difference might partly be due to the fact that the $b_c$-values are different between in 2D and 1D, {\it i.e.\/}, $b_c=37$ in 2D {\it vs\/}. $b_c=37$ in 1D, so that the same $b$-values may effectively have different meaning when scaled by the corresponding $b_c$-value: here $L_{r,max}$ is an increasing function of $b$ in the range of $b$-values studied. The other reason, perhaps more important one, might be that, in 2D, the rupture-length distribution in the region III exhibits a clear downward behavior which serves to shorten the $L_{r,max}$-value. Since the region III corresponds to larger events where the anisotropy of the rupture propagation explained above becomes most eminent, the underlying difference in geometry of 1D and 2D is likely to be playing playing a role here.

 In order to highlight the difference in the behaviors of the event-size distribution in each size region in 2D, {\it i.e.\/}, the regions I$\sim$III, we show in Fig.7 (a) the rupture-length distribution and (b) the magnitude distribution for the representative case of $b=60$, in which the data of the regions I, II and III are separately indicated.  

\begin{figure}[ht]
\begin{center}
\includegraphics[width=7cm]{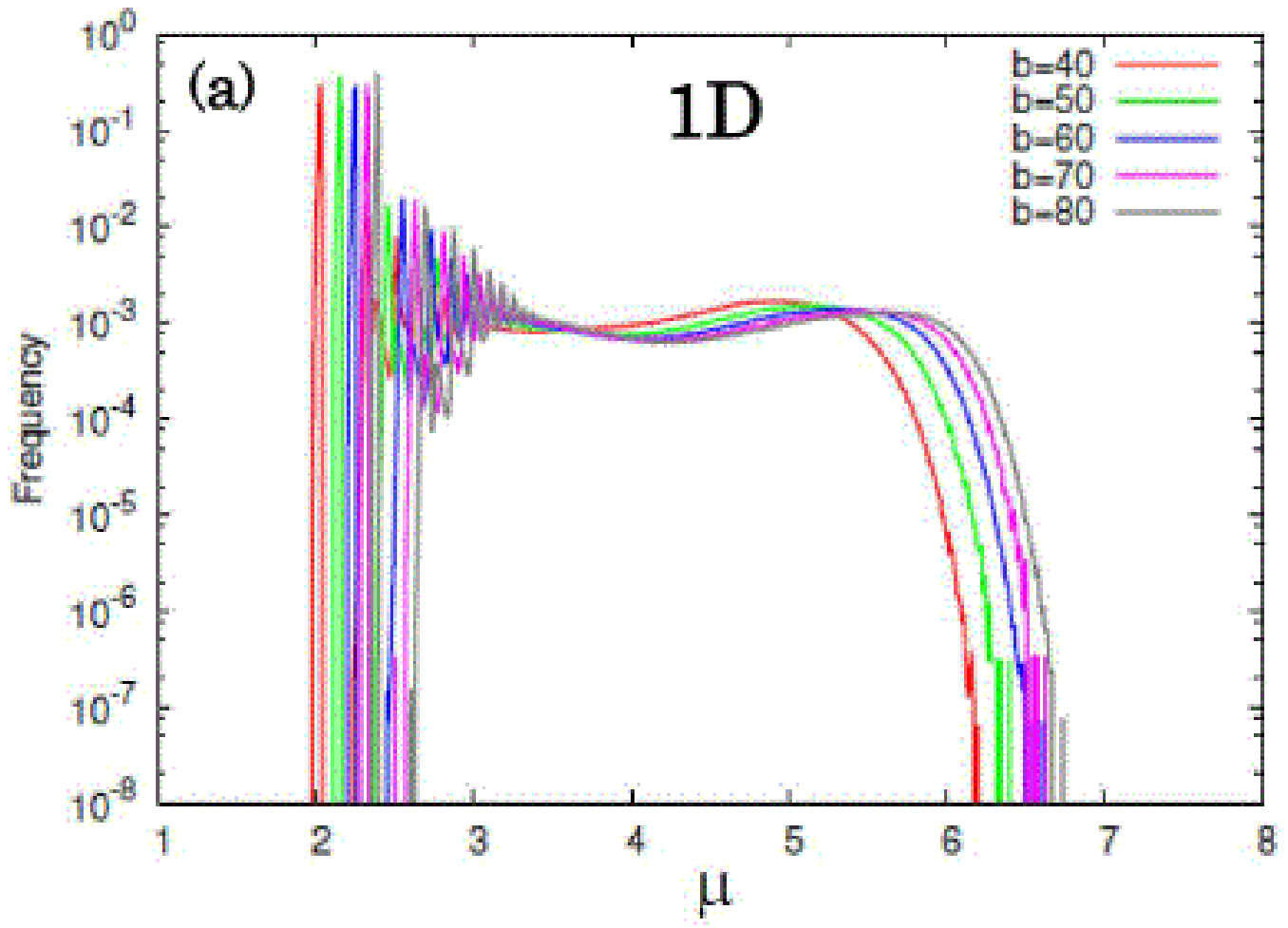}
\includegraphics[width=7cm]{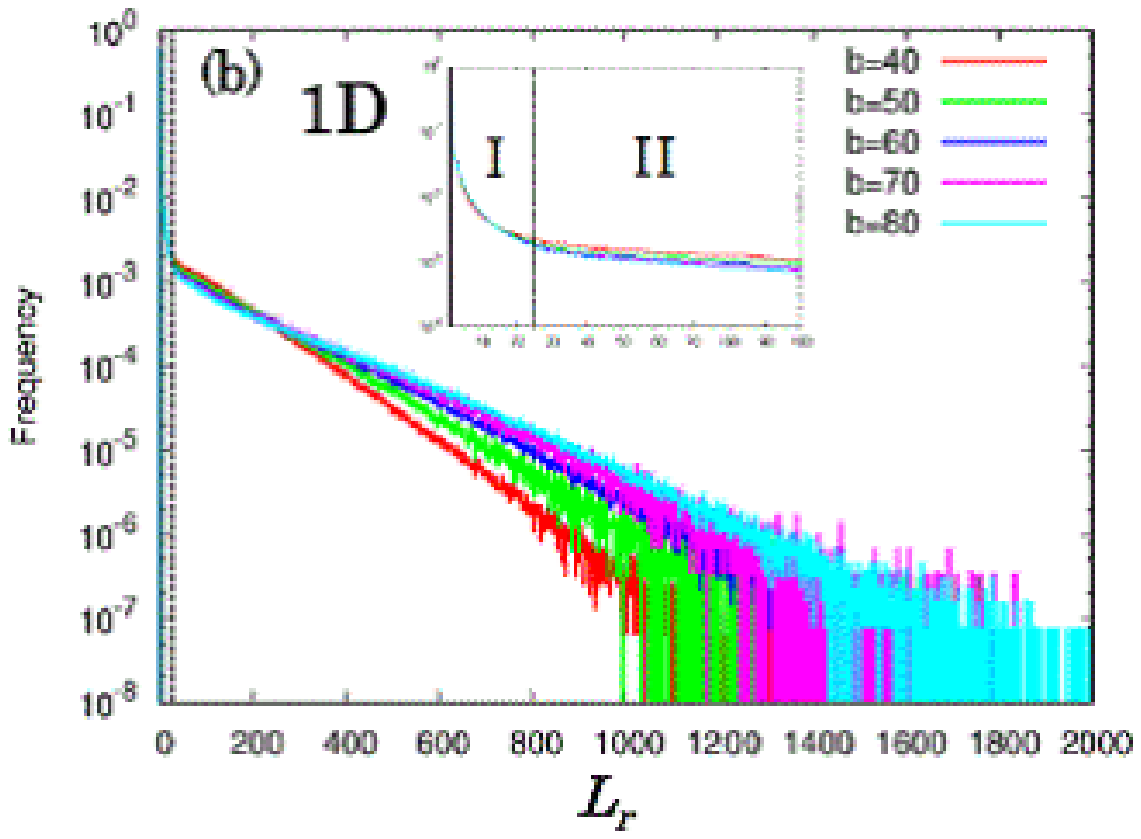}
\end{center}
\caption{
The event-size distributions of the 1D BK model under the RSF law for various values of $b$ in the strong frictional instability regime; (a) the magnitude distribution, and (b) the rupture-length distribution. The lattice size is $L=6400$. The model parameters are $a=1$, $b=60$, $c=1000$, $\ell=3$ ($b_c=19$), $v^*=10^{-2}$ and $\nu=10^{-8}$. The inset is a magnified view of the smaller-$L_r$ region.
}
\end{figure}
\begin{figure}[ht]
\begin{center}
\includegraphics[width=7cm]{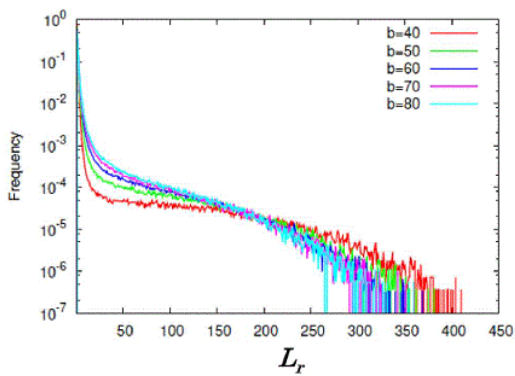}
\end{center}
\caption{
The rupture-length distribution of the 2D BK model under the RSF law for various values of $b$ in the strong frictional instability regime. The lattice size is $L=500$. The other parameters are $a=1$, $c=1000$, $\ell=3$ ($b_c=37$), $v^*=10^{-2}$ and $\nu=10^{-8}$.
}
\end{figure}
\begin{figure}[ht]
\begin{center}
\includegraphics[width=7cm]{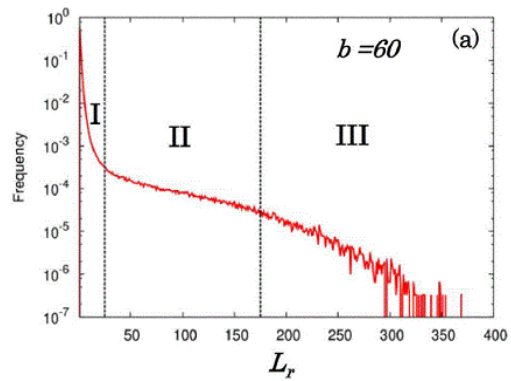}
\includegraphics[width=7cm]{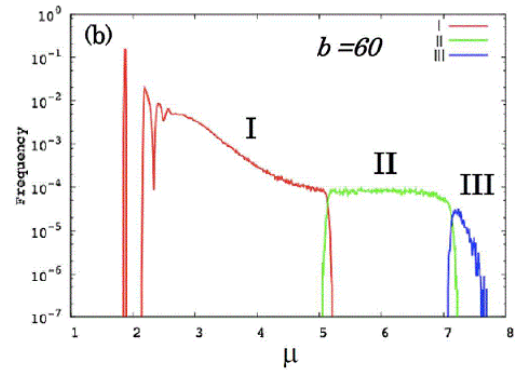}
\end{center}
\caption{
(a) The rupture-length distribution and (b) the magnitude distribution of the 2D BK model under the RSF law for $b=6$, represented separately for smaller events (region I), medium-size events (region II) and larger events (region III).  The lattice size is $L=500$. The other parameters are $a=1$, $c=1000$, $\ell=3$ ($b_c=37$), $v^*=10^{-2}$ and $\nu=10^{-8}$.
}
\end{figure}

As mentioned, in the present 2D BK model, almost all large events are of type-I, in the sense that the epicenter of the event lies next to the rim of the rupture zone of some past event, and the target event tends to be unilateral where the rupture propagates in a direction away from the rupture zone of the past event. Such features are quite natural if one recalls that the block next to the rim of the rupture zone of some past event tends to be in the higher-stress state, which might trigger the subsequent event as an epicenter. Meanwhile, the blocks in the interior of the rupture zone of that past event tends to be in the lower-stress state simply because the stress has already been released by that past event. Then, the rupture of the subsequent event starting from the epicenter block in the higher stress state located next to the rim of the rupture zone of the past event propagates avoiding the rupture zone of the past event in the lower stress state.

 On the other hand, the type-II events where the epicenter is located in the interior of some past event turns out to be rather rare in 2D, at least as compared with the 1D case. In fact, this has been anticipated in Ref.\cite{Kawamura2017} from the geometry of the 2D model and from the stabilization mechanism of the type-II events observed in the 1D model. Our present calculation of the 2D model has confirmed the conjecture of Ref.\cite{Kawamura2017}.

 For the type-I events, we can define the interval or recurrence time $T$ as the time elapsed between the target event and the past event which the target event derived from. In Fig.8, we show the recurrence-time $T$ distribution for various values of $b$ in the strong frictional instability regime, in which the time is normalized as $\nu T$. As can be seen from the figure, the distribution shows a single peak at around $\nu T\simeq 150-200$, which roughly corresponds in our unit to several hundreds years. A clear peak structure suggests the existence of a characteristic time scale for the recurrence of large events, and is consistent with the characteristic feature observed in the size distribution of larger events. An interesting observation is that the recurrence-time distribution also possesses a nonzero weight at $T=0$, which is especially pronounced for larger $b$. The event at $T\simeq 0$ means a twin event. Hence, although there exists a characteristic recurrence time of large events in the present model, a twin-like event which occurs immediately after the first event is also quite possible. It means that, in spite of the existence of the characteristic time scale for the recurrence of large events, a trustable prediction concerning the timing of the next event is still a difficult task even within this simple model. 

\begin{figure}[ht]
\begin{center}
\includegraphics[width=7cm]{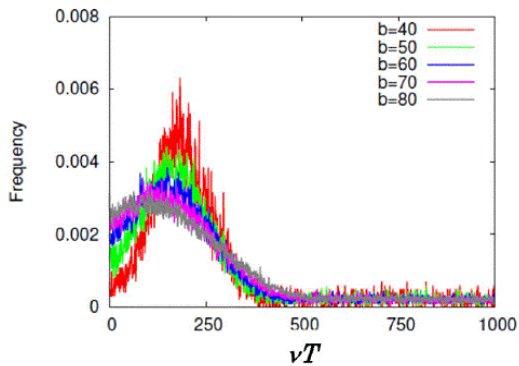}
\end{center}
\caption{
The recurrence-time $T$ distribution of the type-I events of the 2D BK model under the RSF law for various values of $b$ in the strong frictional instability regime. The recurrence time is normalized by the plate velocity $\nu=10^{-8}$. For the definition of the type-I event and the associated recurrence time, see the main text. The lattice size is $L=500$. The other parameters are $a=1$, $c=1000$, $\ell=3$ ($b_c=37$), $v^*=10^{-2}$ and $\nu=10^{-8}$.
}
\end{figure}

 We also investigate the temporal variation of the event frequency just before and after main shocks, paying attention if aftershocks or foreshocks obeying the Omori law or the inverse Omori law are to be realized in the present model. We find that such aftershock (foreshock) sequence obeying the Omori (inverse Omori) law is not observed in the present model, as reported in the corresponding 1D model.

\section{Summary and discussion}

 We studied by extensive computer simulations the nature of high-speed ruptures or main shocks of the 2D BK model obeying the RSF law. Comparison is made with the properties of the corresponding 1D model. We have found that in 2D the rupture propagation on longer length scales is highly anisotropic and irregular in shape, although the rupture-propagation velocity is kept constant $\simeq \ell$ everywhere at the rupture front. Obviously, the nontrivial geometry of the two-dimensionality makes such anisotropic and irregular rupture propagation ever possible. The manner of the rupture propagation sometimes mimics the succesive ruptures of neighboring ``asperities'' occasionally observed in real large earthquakes.
 It should be stressed that the model itself is completely uniform here. The anisotropic and inhomogeneous behaviors of the rupture of the model is dynamically self-generated in its stress state via the completely homogenous evolution rule and uniform material parameters. In seismology, when eminent inhomogeneity and irregularity are observed, they are usually ascribed to the inhomogeneity of the material parameters. Our present result, however, gives a caveat to such a naive interpretation. Even a completely uniform system could, at least in principle, exhibit a strongly inhomogeneous and irregular behavior in its rupture. Of course, there also exists a good possibility that the inhomogeneity in the usual sense, {\it e.g.\/}, inhomogeneity in material parameters, geography of the fault, water contents {\it etc\/.} plays a crucial role. In any case, the true physical origin of the so-called asperities needs further careful examination.

 Large events of the present 2D model tend to be unilateral, with its epicenter lying at the rim of the rupture zone. In fact, the epicenter of the target event is located just next to the rim of the rupture zone of some past event, the rupture propagating away from the rupture zone of the past event. Hence, the events of the 2D model are of type-I in the terminology of Ref.\cite{Kawamura2017}. In the recurrence of large events, while there is a characteristic recurrence time, an immediate recurrence of large events in the form of twin events is also possible. 

 Event-size distributions such as the magnitude distribution and the rupture-length distribution were also studied. The event-size distributions of the 2D model exhibit behaviors a bit different from those in 1D. It exhibits a GR-like power-law behavior for smaller magnitudes, which changes over to a more characteristic behavior for larger magnitudes. In fact, such a change of behavior, {\it i.e.\/}, the GR law holding for smaller magnitudes but a significant deviation from the GR law at larger magnitudes has been reported also in real seismicity \cite{Davison,Wesnousky,Ishibe}.

 The computed rupture-length distribution also exhibits some difference between in 1D and 2D. In particular, the maximum rupture length $L_{r,max}$ is considerably longer in 1D than in 2D. Indeed, while $L_{r,max}\simeq 300-400$ in 2D, it reaches $L_{r,max}\simeq 1000-1500$ in 1D. Presumably, the nontrivial geometry associated with the rupture propagation in large events in 2D causes such a difference.


 We hope that our present analysis of the uniform BK model in 2D under the RSF law might be of some help in understanding the complex nature of earthquakes.

 This study is supported by a Grant-in-Aid for Scientific Research No.16K13851, and by the Ministry of Education, Culture, Sports, Science and Technology (MEXT) of Japan, under its Earthquake and Volcano Hazards Observation and Research Program.

\end{document}